# High-resolution wide-field OCT angiography with a self-navigation method to correct microsaccades and blinks

**XIANG WEI,**[1,2] **TRISTAN T. HORMEL,**[1] **YUKUN GUO,**[1] **THOMAS S. HWANG,**[1] **AND YALI JIA**[1,2*]

[1]*Casey Eye Institute, Oregon Health & Science University, Portland, Oregon, 97239, USA*
[2]*Department of Biomedical Engineer, Oregon Health & Science University, Portland, Oregon, 97239, USA*
*\*jiaya@ohsu.edu*

**Abstract:** In this study, we demonstrate a novel self-navigated motion correction method that suppresses eye motion and blinking artifacts on wide-field optical coherence tomographic angiography (OCTA) without requiring any hardware modification. Highly efficient GPU-based, real-time OCTA image acquisition and processing software was developed to detect eye motion artifacts. The algorithm includes an instantaneous motion index that evaluates the strength of motion artifact on *en face* OCTA images. Areas with suprathreshold motion and eye blinking artifacts are automatically rescanned in real-time. Both healthy eyes and eyes with diabetic retinopathy were imaged, and the self-navigated motion correction performance was demonstrated.

## 1. Introduction

Optical coherence tomography (OCT) is a non-invasive modality capable of detailed imaging of retinal and choroidal structure [1]. OCT can also produce angiographic data (OCTA) by measuring motion contrast between successive OCT images [2-5]. Compared to conventional angiographic imaging techniques such as fluorescein angiography (FA), OCTA provides superior resolution as well as volumetric data [6]. As a non-invasive imaging method, OCTA also avoids the potential side effects and discomfort associated with dye injection [6].

Wide-field OCTA imaging has recently attracted significant research interest. Some diseases, like diabetic retinopathy (DR), can manifest with predominantly peripheral vascular changes not visible in typical OCTA macular scans [7]. While it has a wider field-of-view (FOV) than OCTA, because of potential side effects FA is not considered suitable for routine imaging [8]. In contrast, wide-field OCT which is faster, safer, and more comfortable for the patient could be used in routine settings. Wide-field OCT has been implemented in studies since 2010, but it has been limited by lower resolution [9]. With improvement in laser sweep-rate and image processing techniques, several groups have explored high-resolution wide-field OCTA [10-12]. To maintain comparable image resolution with a larger FOV, the total number of sampling points needs to be greatly increased along both the fast and slow axes. Typically, the sampling density should meet the Nyquist criterion. This results in longer inter-frame and total imaging times, which in turn exacerbates artifacts due to microsaccadic motion, blinking, and tear film evaporation [13]. Hardware improvements can yield high-resolution, large FOV OCTA simply by increasing scan rate [11, 14], and many groups have developed such fast-scanning systems [14-17]. Still, even in state-of-the-art systems, data acquisition requires some trade-off between image resolution and FOV size [11].

Successful widefield systems, therefore, require some means of artifact minimization. One widely used approach especially on laboratory devices is postprocessing [18-23]. Postprocessing algorithms can be used to suppress certain artifacts [24]. However, postprocessing corrections involve some degree of information loss [18] and may require multiple volume acquisitions, thereby increasing the total imaging time [18]. Multi-image registration and montage can further enlarge the FOV of OCTA images [25, 26]. However, image stitching can introduce new artifacts while simultaneously increasing acquisition time and difficulty [27].

An alternative approach to artifact minimization is to leverage real-time tracking system [28]. Some prototypes [28-31] and most commercial OCT systems such as RTVue (Optovue Inc., USA), CIRRUS (Carl Zeiss AG, Germany) and Spectralis (Heidelberg Engineering, Germany) have already incorporated motion tracking. They require additional imaging hardware in the OCT optical axis to simultaneously acquire images. Infrared fundus photography or scanning laser ophthalmoscopy (SLO) [28-30] are two common allied methods used to perform tracking. Such schemes have several drawbacks. Both fundus camera and SLO images need to be acquired and processed separately from the OCT image, increasing the processing complexity of the system [28]. In addition, SLO and fundus photography can only provide indirect information about the OCT image quality. Furthermore, additional hardware can increase the cost of OCT system and make maintenance and repair more difficult.

Apparently, a method that does not require additional hardware can obviate these problems [32]. One possible solution is a blink and motion artifact correction method directly based on OCTA, which often already calculates decorrelation [3] that can be used to detect motion. Here, we introduce a novel OCTA-based self-navigated motion correction method that can efficiently remove blink and motion artifacts without additional hardware support. We use the term "self-navigated motion correction" in order to emphasize that our system automatically detects motion, but relies on the imaging subject to fixate on a target in order to correct this motion when it is detected. This is unlike conventional active tracking methods used in prototypes and commercial systems, since the self-navigated motion correction reported here is a passive technology that does not actively track eye movements. As we will show, this self-navigated motion correction technique enables rapid acquisition of wide-field OCTA data while minimizing motion artifacts.

## 2. Method

### 2.1 400-kHz swept-source OCT (SS-OCT) system

This system uses a customized 400-kHz swept-source laser (Axsun Technologies, USA), which is 4-6 times faster than the laser used in commercially available devices. The laser has a center wavelength of 1060 nm with 100 nm sweep range operating at 100% duty cycle. The maximum theoretical axial resolution is 4 µm in tissue [33]. With an appropriate optical design of the sample arm we recently reported [11] that a 75-degree maximum FOV can be achieved. Different from our previously study, the spot size on the retina was reduced to 10 µm, which is equivalent to the maximum lateral resolution that this system can achieve.

### 2.2 GPU-based real-time OCT/OCTA

Graphics processing unit (GPU)-based parallel computing techniques can significantly improve the OCT and OCTA data processing speed. Many groups have developed GPU-based data processing techniques previously [34-36]; with the help of GPU parallel processing, a megahertz A-line rate data processing speed can be achieved [37]. Real-time OCTA image processing has also been realized previously [38-40]. In this study, we developed GPU-based real-time OCT/OCTA data acquisition and processing software for swept-source OCT system modified from our previous work [40]. The split-spectrum amplitude-decorrelation angiography (SSADA) algorithm was applied to compute OCTA flow signal [3]. SSADA increases the flow signal-to-noise ratio by combing flow information from each split-spectrum. Different from previous study, here, the GPU-based parallel data processing further improves real-time efficiency by processing the OCT and OCTA images together in a single GPU thread (Fig. 1).

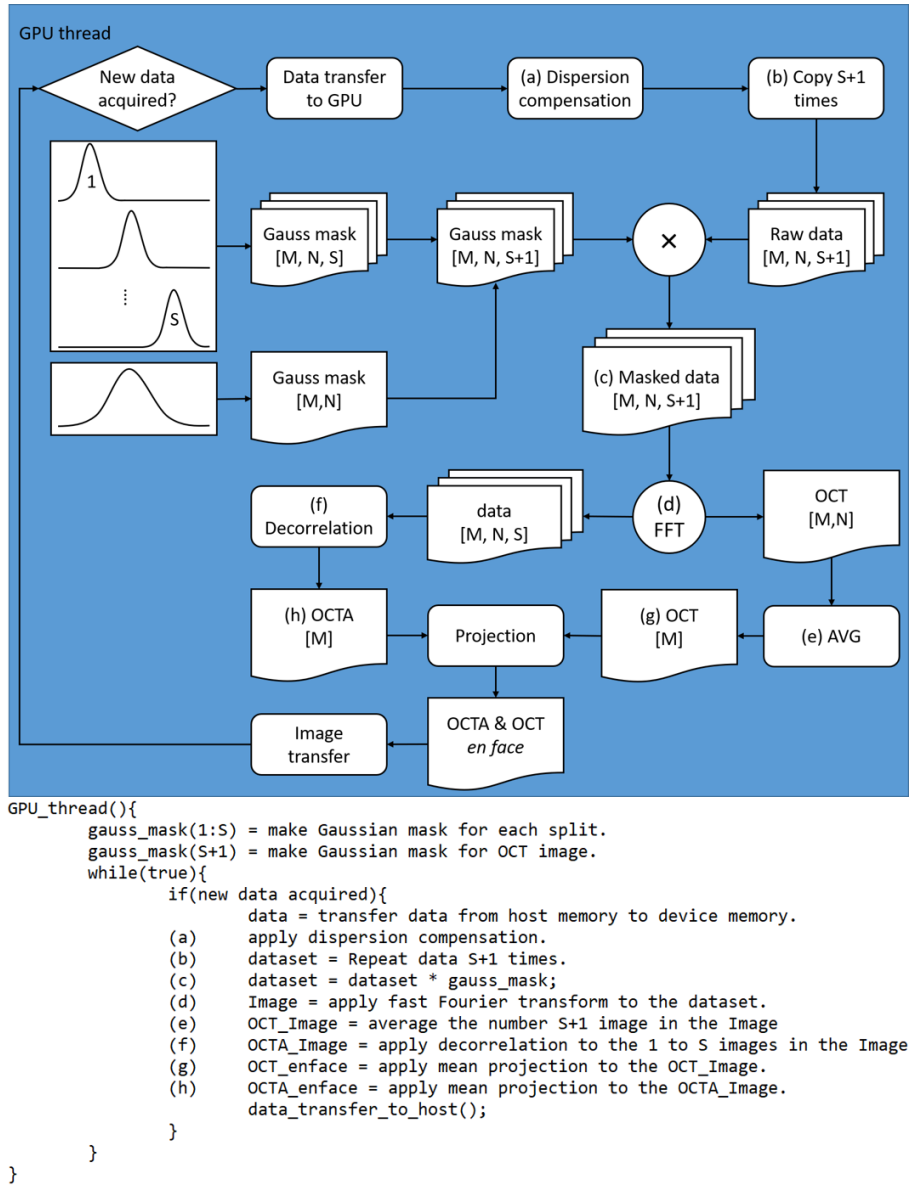

Figure 1. GPU-based real-time OCT/OCTA image processing thread. Once the raw spectrum batch is acquired and transferred from the host memory to the device memory, the thread will begin to process the new data. The dataset contains spectra from M location with N repeated scans at each location. A GPU-based dispersion compensation algorithm is applied to the raw spectrum (a). The raw spectrum is then copied to form a dataset with S+1 copies, where S indicates the number of spectrum splits, and the one additional copy is used to generate the OCT image (b). Then a set of Gaussian masks are applied to the dataset (c). A fast Fourier transform (FFT) is then applied to the combined dataset, which generates a split spectrum image and OCT image (d). The OCT data is then averaged across N repeats to generate an averaged OCT image, which reduces noise (e). The split spectrum image set is then processed using a decorrelation algorithm to generate the OCTA image (f). Then the OCTA and OCT images are projected using mean projection to generate an en face image (g, h). The following pseudo code is used to demonstrate the process.

Instead of processing each B-scan separately, a batch of B-scans is processed together. First, the raw spectra are processed to compensate for dispersion. Then, a pre-generated Gaussian mask matrix is applied. This matrix contains 11 narrow Gaussian masks for OCTA image processing and one additional broad Gaussian mask for structural OCT data processing. For each processing batch, the fast Fourier transform is only applied once to minimize the processing time. After the fast Fourier transform, the OCT image is generated by averaging the repeated scans. The remaining 11 OCT images from the spectrum splits are further processed using amplitude decorrelation algorithm. The speed bottleneck of real-time processing is not the data computation, but the data transfer. To avoid extra time wasted beyond the data acquisition time, 12 B-scans at four locations were batched for each processing unit. The total data computation and transfer time for each batch is less than 30 ms, lower than the acquisition time of 42 ms for 12 B-scans. The mean values of OCT and OCTA cross-sectional images are projected to generate OCT and OCTA *en face* images. Both cross-sectional and *en face* images are displayed on a custom graphical user interface (GUI) in real-time.

## 2.3 Self-navigated motion correction OCT/OCTA

Conventional real-time motion detection algorithms are based on the correlation between two sequentially acquired fundus *en face* images either from infrared camera or SLO. The information used to calculate correlation is mainly from the major retinal vasculature. This procedure requires additional hardware and software support. OCTA generates vasculature and motion signal itself; therefore, its cross-sectional frames (B-frames) are inherently co-registered with the location of eye blinks and motions. To better represent the motion strength quantitatively, we defined an instantaneous motion strength index (IMSI) using the normalized standard deviation of the *en face* OCTA data:

$$\text{IMSI} = \frac{\text{std}(D_{\text{OCTA}})}{\text{mean}(D_{\text{OCTA}})} \tag{1}$$

The IMSI was calculated in a single CPU thread after the OCTA *en face* image is generated in the GPU. $D_{\text{OCTA}}$ represents the mean projection of OCTA values from single batch (4 OCTA B-frames generated from 12 OCT B-scans). A fixed threshold is set to highlight all involuntary motion that is intensive enough to affect image quality. The performance of this motion index was evaluated using wide-field *en face* OCTA images generated in real-time from a healthy human volunteer (Fig. 2).

Besides the detection of microsaccades, blink detection is also required to achieve artifact-free OCTA imaging. During high-resolution wide-field OCTA image acquisition with our system, the imaging subject can freely blink several times in order to keep the tear film intact. During the blinking time course, the signal strength is significantly reduced. One straightforward approach for blink detection is to set a secondary threshold on the signal strength of the OCT structural image. When the signal strength is subthreshold, a blink may be detected. However, this approach may cause many false detections during wide-field imaging. A major artifact source in wide-field OCT image is shadow artifacts caused by vignetting and vitreous opacity, which frequently occur in wide FOV imaging and significantly reduce signal strength. In this old-fashioned way, the signal strength decrease caused by shadow artifacts can be mistakenly detected as a blink. However, at the initiation of a blink, the eye usually has a large axial movement, which can be detected by IMSI. When the eye is closed, the OCTA signal has low variation, and the IMSI value is low across an entire batch. The IMSI calculated from combining the batch at the blink initiation and eye closure will yield a high value, indicating blinking artifacts. In our motion and blink detection mechanism, the IMSI is a normalized metric, independent of the variation of OCTA signal strength. This single fixed IMSI threshold can be used to handle blinks and various types of motions across different imaging subjects and different systems.

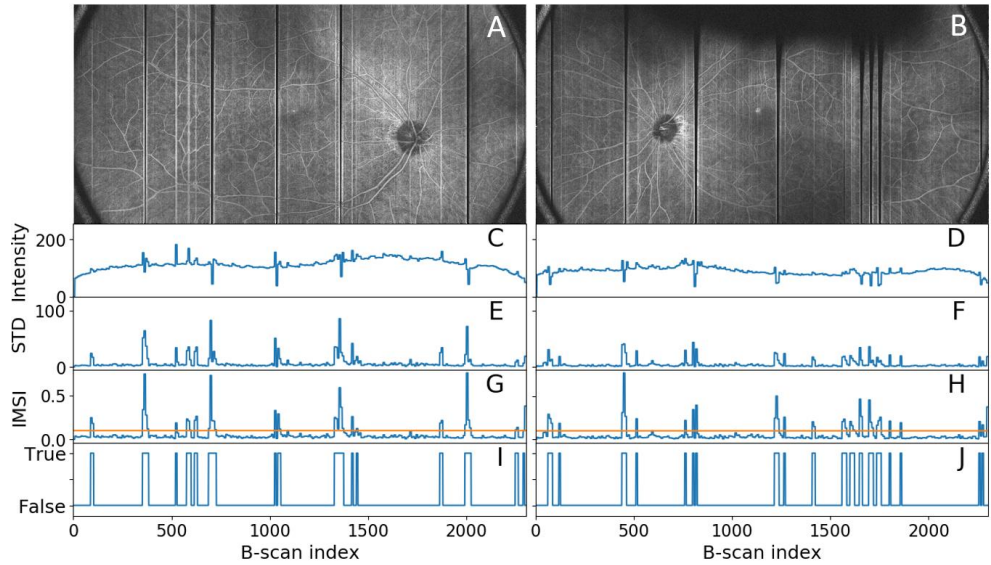

Figure 2. (A, B) *En face* OCTA images acquired from a healthy volunteer without the motion correction system engaged; (C, D) mean values from each OCTA batch; (E, F) standard deviation (STD) between B-frames in each batch; (G, H) IMSI calculated using the *en face* OCTA image, with yellow line indicating the threshold; (I, J) motion trigger signal generated after the threshold applied to the IMSI value.

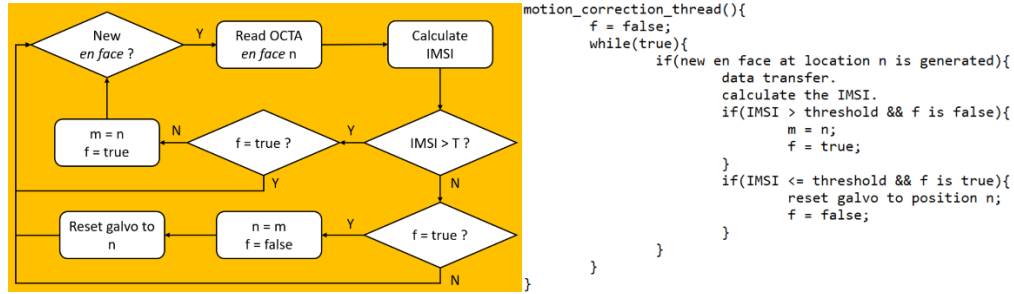

Figure 3. Flow chart of the GPU-based real-time self-navigated motion correction thread. Variable n indicates the global batch number counter, variable m indicates a batch that contains artifacts, and f is the motion flag. After the GPU thread generates a new *en face* image, the image is then processed to calculate IMSI to determine if motion occurred. If motion occurred, the thread will wait until no additional motion events occur to reset the galvo scanner. The following pseudo code is used to better demonstrate the process.

After accurately detecting motion or blink artifacts, the system can then automatically rescan the artifact affected area to restore the image quality (Fig. 3). When eye blinks or microsaccades are detected, the system will record the B-scan number when the motion begins, it will continue scanning and acquiring the following positions. The following batches' IMSI will also be calculated. When the IMSI values return to the normal level (subthreshold), the system will reset the scanner and calculate the IMSI from the new batch acquired after the scanning reset. By doing so, the image can be re-evaluated. If additional artifacts are detected at the same position, the scanner will reset again to rescan the artifact affected area until the blink or motion process is completed. As can be seen (Fig. 4), this procedure is capable of correctly removing blink artifacts.

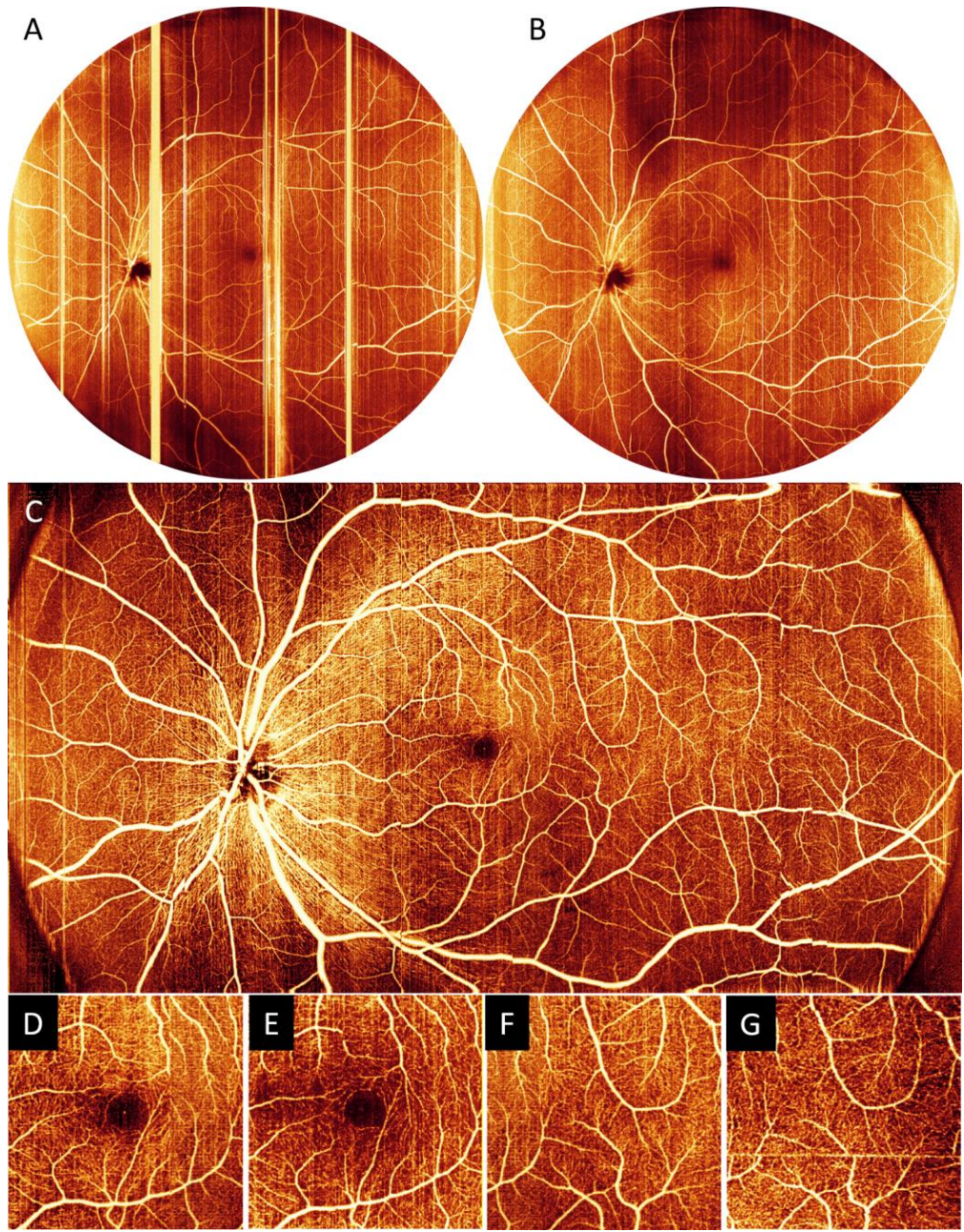

Figure 4. High-density wide-field OCTA images. (A) wide-field OCTA image acquired without self-navigated motion correction engaged; (B) wide-field OCTA image acquired with self-navigated motion correction engaged. The shadowing artifact is caused by eye lashes; (C) high-resolution wide-field OCTA image acquired with self-navigated motion correction engaged; (D, F) 3×3-mm inner retinal angiogram cropped from C; (E, G) 3×3-mm inner retinal angiogram acquired using commercial system for comparison.

## 3. Results

*3.1 Healthy retina*

To evaluate performance of the self-navigated motion correction system, we applied our algorithm with two different scanning patterns on a healthy volunteer. First, we acquired two wide-field high-resolution OCTA images from a healthy human volunteer, once without (Fig. 4A) and once with (Fig. 4B) the self-navigated motion correction system. For each volume, images contained 2560 A-lines per B-scan and 1920 B-scans per volume, with two repeats. A bidirectional scan pattern [41] was applied here. The sampling step size in the fast axis was 9 µm and in the slow axis was 12 µm. The total data acquisition time was 25 seconds without self-navigated motion correction. In this work, the total scanning time of the image with self-navigated motion correction was always less than 1 minute. During the data acquisition with real-time motion correction engaged, when the volunteer blinked the scanned eye, the system successfully detected each blink and rescanned the area right after the eye re-opened. Large artifacts caused by microsaccadic movements were also successfully detected and the affected areas can be rescanned right after the motion processes were completed. At the edge of the rescan waveform, there are fly-back artifacts, but the artifacts are only limited to 50 A-scans which is only 1/20 of the total A-scan in each B-scan. In order to complete the scan within a reasonable time (<1 min), very mild motion artifacts that were lower than the IMSI threshold were deliberately disregarded. After inner retinal layer was segmented [42], its angiogram was generated by projecting the maximum OCTA value within this slab [43]. A Gabor filter, histogram equalization, and a custom color map were applied to enhance the visualization.

A high-density retinal OCTA scan with horizontal 75-degree and vertical 38-degree FOV (23 ×12- mm) was acquired on the same healthy eye. A raster scan pattern was applied here. The scan contained 1208 A-lines per B-scan, 2304 positions per volume with 3 repeated B-scans at each position, which achieves the sampling density of 10 µm/line on both horizontal and vertical axes. Our results demonstrated that the *en face* OCTA (Fig. 4C) is free of blink and large motion artifacts, although a few minor artifacts still remain; the high sampling density enabled acquisition of high-resolution angiograms with fine vascular details. To further validate its image quality, the same healthy human subject was also scanned using a commercial OCT system (RTVue-XR Avanti; Optovue, Inc., Fremont, CA), with $3 \times 3$-mm ($304 \times 304$ lines) retinal images acquired in both the central macular and the peripheral temporal regions. For a fair comparison, only a single OCTA scan volume from RTVue-XR were used to generate the inner retinal angiogram. The cropped images from high-resolution wide-field prototype OCTA at the same position were used for side-by-side comparison (Fig. 4C). For a qualitative inspection, the images from the prototype system and the commercial system show similar image quality and capillary visibility (Fig. 4D, E, F, G).

We also evaluated the performance of the self-navigated motion correction system quantitatively. We scanned 14 eyes from 7 healthy human subjects with and without the motion correction and examined the real-time OCTA *en face* images generated along with the OCT spectrum raw data. Without the motion correction system, 104 blinks were present across the images. With the self-navigated motion correction system, no blink artifacts were present in the images. The motion was measured automatically by calculating the IMSI and detected by setting a threshold at 0.25. Without motion correction, 1,976 significant movements were detected in total. The number of artifacts was dropped to 168 when self-navigated motion correction was engaged; and the blink and motion artifacts were suppressed by 100% and 91.5% respectively.

*3.2 Retina with diabetic retinopathy*

A $23 \times 12$-mm OCTA using the same scan pattern described above was acquired from a 57-year-old female diagnosed with proliferative DR and early stage cataract. Our prototype system does not require dilation, but to minimize the difficulty of alignment during imaging this participant was dilated. The resulting OCTA demonstrated capillary dropout, microaneurysms, and intraretinal microvascular abnormalities in high resolution with a FOV that exceeds the current commercial OCT systems (Fig. 5).

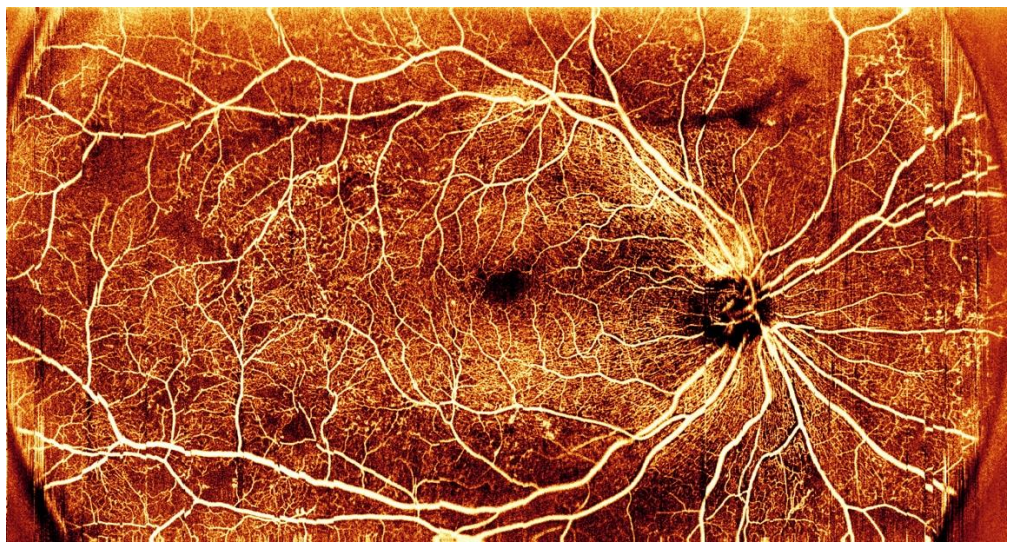

Figure 5. Inner retinal angiogram acquired by our prototype high-resolution wide-field OCTA from a participant with proliferative diabetic retinopathy.

## 4. Discussion

In this paper, we presented a novel OCTA-based microsaccade and blink detection and correction method that does not require additional hardware or modification to an OCT/OCTA device. We applied this method to image both healthy and DR eyes, acquiring 75-degree wide-field capillary-level resolution OCTA with comparable capillary visibility to $3 \times 3$-mm commercial OCTA images. To our knowledge, this is the first application of a real-time self-navigated motion correction system on an OCTA device that relies exclusively on OCTA data for motion artifact correction. Previously, OCT systems either used infrared fundus photography or infrared SLO as the reference for eye motion and blink detection. Our system has several advantages over such conventional tracking and motion correction methods. First, since no additional hardware is required, it reduces the complexity and cost of OCT systems; second, OCTA provides richer vasculature information to evaluate motion amplitude; third, fundus images are not adequately sensitive to small changes in vessel location to precisely align OCTA images due to their relatively low resolution. OCTA is intrinsically sensitive to movements (such as microsaccades) which make the detection of motion much easier than third party tracking methods.

GPU-based data processing technology accelerated the development of real-time OCT/OCTA. Our self-navigated motion correction method relies on high-speed, real-time OCTA image processing to respond to microsaccade and blink artifacts. The current bottleneck in real-time OCTA image processing is data transfer speed. For real-time applications, the computation and transfer time needs to be less than the data acquisition time. For our 400-kHz OCT system with 1204 A-lines per B-scan, a minimum of 12 B-scans should be processed together in batches in a Nvidia RTX 2080ti GPU to maintain real-time processing. A larger batch can increase data computation and transfer efficiency at the cost of increased data transfer time. The motion correction system must balance the processing efficiency and the correction response time. A large batch will have a longer acquisition and processing time, which will result in a long response time. A slow response correction system will waste many "good" B-frames and directly extend the data acquisition time.

In our self-navigated motion correction system, one of the most important features is the IMSI metric. As the key motion indicator, the IMSI reliability directly correlates to the reliability of the motion correction. Here, IMSI is a normalized value across several different B-scans. The normalization process removed the dependency on signal strength, yielding a pure correlation to motion. Thus, in our self-navigated motion correction algorithm, IMSI is independent to variation between different types of motions, different imaging subjects, different SNR, and different types of systems. However, IMSI is still affected by the number of B-frames used in each batch. In our experience, if the processing batch is too small, it renders an unreliable IMSI. In our system, we used a high-speed swept-source laser, which requires at least 4 OCTA B-frames in each batch.

Wide-field OCTA imaging is challenging, and ocular pathology and involuntary motion can reduce image quality. Increasing the speed and sensitivity of the system can possibly overcome some of difficulties. Many groups have applied high-speed swept-source lasers [14, 44] to prototype OCT systems. One of the fastest swept wavelength laser sources is the Fourier domain mode lock (FDML) laser. It can achieve megahertz swept source rates. At such a high speed, video-rate OCT imaging has been realized [45]. In our system, we employed a 400-kHz rather than a megahertz swept source laser. There are several considerations when selecting a laser source. One is that increasing sweep rate decreases the SNR of OCTA system. Another is the scanning speed. High scanning speeds require a resonant scanner, which frequently causes image distortion problems. Furthermore, high B-scan rates reduce the flow SNR and OCTA image quality due to diminished vascular flow contrast. Finally, for wide-field imaging, a sufficient imaging range is required. For a 3-megahertz OCT system, suppose we need 1536 pixels per A-line to achieve similar imaging depth in our system. To acquire those pixels, a 5 GHz balanced detector would be needed. However, currently it is a challenge to design such a balanced detector. We opted for the 400-kHz source for these reasons.

Another challenge for wide-field OCTA imaging is shadows. In our previous study, a special optical system was designed to eliminate the shadow caused by pupil misalignment [11]. This system was also employed in this work. The cross-scanning pattern applied in preview step can also increase the alignment efficiency. Compared to the conventional multi-position method, the method presented here has a higher refresh rate that can show shadows in real-time. Doing so can help control scan quality.

Our current motion correction system still has some limitations. The system can only provide an indicator that microsaccades and blinks occurred, and does not provide any quantitative lateral or axial motion trajectory information. Without such information, the system is reliant on the fixation target and the cooperation of the patient to realign the eye after movement, which may introduce artifacts like vessel interruption. Lack of lateral motion information can also cause low sensitivity to slow motion (like ocular drift). Even with a small fixation target, the change of gaze after each motion or blinks can also introduce vessel interruptions and some information loss. Evaporation of tear film during the imaging session will also reduce the image quality, as this leads to loss of focus. Several blinks are required to recover the tear film, which increases the chance of more vessel interruptions and information loss. Because our approach requires the imaging subject's cooperation, this system cannot work on sedated patients, non-human primates and neonates. As a passive system, the image quality also relies on patient stability and ability to concentrate on the fixation target. An imaging subject with difficulty focusing on the fixation target and keeping steady during the imaging process will decrease the image quality.

Finally, we took advantage of GPU and SSADA parallel processing efficiency. Together, these software and hardware improvements enabled the high-quality images presented in this work.

## 5. Conclusion

We have successfully developed a novel real-time OCTA-based microsaccade and blink detection and correction system for high-resolution wide-field OCT/OCTA imaging. This motion correction system is integrated in a high-speed swept-source OCT system capable of acquiring a 75-degree field-of-view high-density OCTA image in healthy and DR patients. By calculating an instantaneous motion index and rescanning in real time, the system detects and eliminates artifacts due to eye blinking and large movements. This system represents a significant improvement in field-of-view and resolution compared to a conventional OCT system.

**Funding**

National Institutes of Health (R01 EY027833, R01 EY024544, P30 EY010572); unrestricted departmental funding grant and William & Mary Greve Special Scholar Award from Research to Prevent Blindness (New York, NY).

**Disclosures**

Oregon Health & Science University (OHSU) and Yali Jia have a significant financial interest in Optovue, Inc. These potential conflicts of interest have been reviewed and managed by OHSU